\documentclass[journal]{IEEEtran}
\usepackage{amsmath,color,graphicx,amssymb}
\usepackage{ifthen}
\usepackage{placeins}
\usepackage{float}
\usepackage[bookmarks=false]{hyperref}
\usepackage{tabularx}
\usepackage{epstopdf,multirow}
\usepackage{xcolor}

\begin{document}
\pagenumbering{gobble}

\title{Surveillance of Space using Passive Radar and the Murchison Widefield Array}
\author{\IEEEauthorblockN{James~E.~Palmer\IEEEauthorrefmark{1},~\IEEEmembership{Senior Member,~IEEE},
Brendan~Hennessy\IEEEauthorrefmark{1},
Mark~Rutten\IEEEauthorrefmark{1},
David~Merrett\IEEEauthorrefmark{1},
Steven~Tingay\IEEEauthorrefmark{2}\IEEEauthorrefmark{4}, 
David~Kaplan\IEEEauthorrefmark{3},
Steven~Tremblay\IEEEauthorrefmark{4}\IEEEauthorrefmark{5},
S.M.~Ord\IEEEauthorrefmark{4},
John~Morgan\IEEEauthorrefmark{4}, and
Randall~B.~Wayth\IEEEauthorrefmark{4}}
\\
\IEEEauthorblockA{\IEEEauthorrefmark{1}Defence Science \& Technology Group, Edinburgh, Australia. 5111} \\
\IEEEauthorblockA{\IEEEauthorrefmark{2}INAF Science Directorate (Divisione Nazionale Abilitante della Radioastronomia). Bologna Italia} \\
\IEEEauthorblockA{\IEEEauthorrefmark{3}University of Wisconsin--Milwaukee, Milwaukee, USA} \\
\IEEEauthorblockA{\IEEEauthorrefmark{4}ICRAR/Curtin University. Bentley, Perth. WA 6102} \\
\IEEEauthorblockA{\IEEEauthorrefmark{5}ARC Centre of Excellence for All-sky Astrophysics (CAASTRO)}}

\maketitle

\begin{abstract}
In this paper we build upon recent work in the radio astronomy community to experimentally demonstrate the viability of passive radar for Space Situational Awareness. Furthermore, we show that the six state parameters of objects in orbit may be measured and used to perform orbit characterisation/estimation.
\end{abstract}


\IEEEpeerreviewmaketitle

\section{Introduction}

\IEEEPARstart{P}{assive radar} exploits readily available, non-cooperative sources of radio frequency (RF) energy as illuminators of opportunity to measure reflections from the environment and objects of interest.  
 
Without the need for deployment and operation of a dedicated transmitter, passive radar systems may be significantly less expensive to implement and operate than their conventional counterparts.  

Surveillance of Space / Space Situational Awareness is an area that is receiving much attention in recent times \cite{NAP13456,7024495} due to the increasing density of objects in orbit and increasing concerns over the so-called Kessler effect \cite{7024495}.  In this paper, we further demonstrate the viability of using FM radio transmitters as the RF donor and a radio astronomy receive array as the surveillance receiver in a passive radar configuration to detect airborne and large space-borne objects of opportunity.

\subsection{Summary of prior work}
The Murchison Widefield Array (MWA) \cite{2013PASA...30....7T} has previously been used in demonstrating detections of the International Space Station and the moon \cite{2013AJ....145...23M, 2013AJ....146..103T} using reflected FM radio emissions.  However, the processing employed in those demonstrations assumed no knowledge of the reference waveform and employed radio astronomy processing techniques to image the sky, resulting in detections in the angular domain (i.e. azimuth and elevation, or more precisely right-ascension, declination) only.  As such, the processing that was applied provided no estimate of the range or velocity of the objects as they migrated through the scene.

In this paper we use passive radar processing techniques to show that the range and velocity information can be obtained for targets whilst also achieving a significant processing gain.

\section{MWA Summary}
\begin{figure}
\includegraphics[width=\columnwidth]{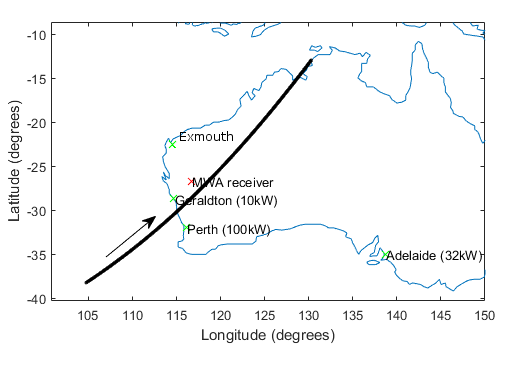}
\caption{MWA Location, Transmitters and approximately two minutes of ISS flight path }
\label{Fig:MWA_loc}
\end{figure}

The Murchison Widefield Array is a radio astronomy pre-cursor development for the Square-Kilometre Array project \cite{5136190}.  The array is located in the Murchison Radio-astronomy Observatory in Western Australia, and its location is shown in Figure \ref{Fig:MWA_loc}.

\subsection{Antenna details}
Some key parameters of the MWA system include:
\begin{itemize}
\item No. of antennas: 128 `tiles' with 2 polarisations (EW, NS)
\item No. of elements per tile: 16 in a 4x4 configuration with analogue beamforming
\item Collection area: approx. 2000 square metres
\item Frequency range: 80 - 300~MHz
\item Instantaneous bandwidth: 30.72~MHz (24 x 1.28~MHz)
\end{itemize}

\begin{figure}
\includegraphics[width=\columnwidth]{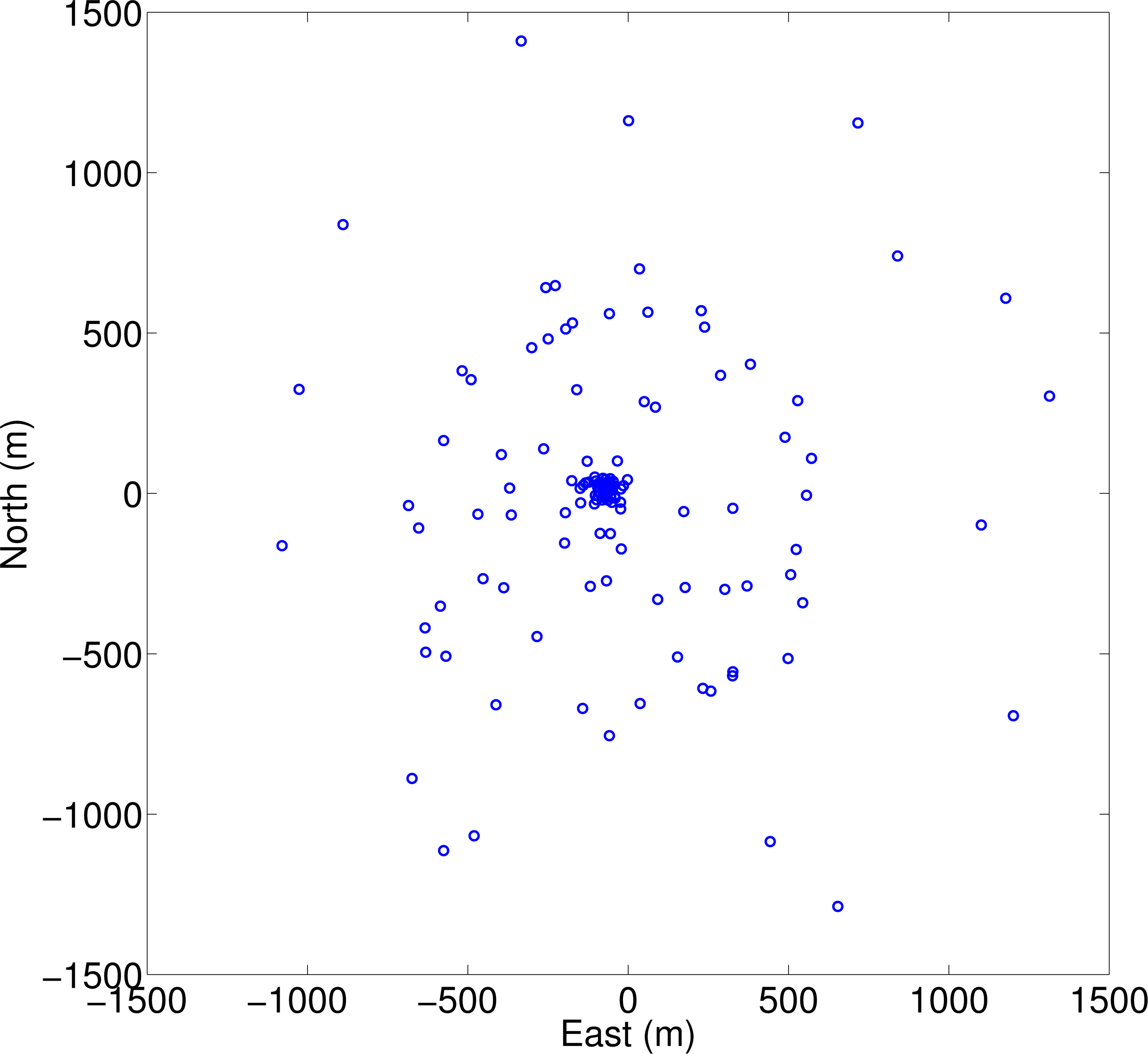}
\caption{MWA Layout - each blue circle represents a tile}
\label{Fig:MWA_layout}
\end{figure}

The array's tiles are laid out in a sparse arrangement with a dense core as shown in Figure \ref{Fig:MWA_layout}.  The overall aperture results in an angular resoution approaching 0.05 degrees.

\section{Processing Strategies}
\label{Sec:proc}
The MWA includes a Voltage Capture System (VCS)\cite{2015PASA...32....5T} for recording high time and frequency resolution data. The output of the analogue beamformers are amplified, digitised and then processed through two polyphase filter bank stages which critically sample the data into coarse 1.28 MHz channels and then 10 kHz fine channels. These data are phase-adjusted to account for the cable delays and Jones matrix-based calibration measurements are also applied in order to remove instrumental and atmospheric effects. These Jones matrices are produced by the MWA system through recursively accounting for residuals after removing the visibilities from known strong compact sources\cite{2013PASA...30....7T}.

The 10 kHz sub-channels are recombined to create the larger bandwidth FM radio signal used in the beamforming and range-Doppler processing. Traditional beamforming, accounting for the phase delays to each of the antenna elements, is used to construct time series data corresponding to very precise directions. Sky visibilities are then calculated from the power in each direction's beam, without having to perform correlation/interferometric processing.

Range-Doppler processing is achieved by receiving two separate channels of data, one being the direct transmitted signal and the other being from the surveillance area of interest. Targets are found by searching for distorted copies of the direct signal in the surveillance signal; by comparing both time of arrival and frequency of arrival differences the target's bistatic-range and relative velocity can be calculated.\cite{4654011}

\section{Experimental Campaign}
\label{sec:exp}
A joint experiment was conducted in April, 2015 using both the MWA and a Defence Science and Technology Group (DST Group) line-of-sight (LOS) receiver deployed in Adelaide, Australia.  The DST Group receiver collected a high SNR reference, whilst the MWA collected the ISS reflections.  Whilst both the LOS receiver and the MWA employ GPS disciplined local oscillators, the collections were started at slightly different (yet overlapping) times.  As will be discussed further in the next section, after processing the data, it became apparent that two other transmitters were also present in the MWA data, namely Geraldton and Perth. Although the original intent of the experiment was to capture emissions from Adelaide's FM transmitter that illuminated the International Space Station (ISS) as it passed over the MWA, the results in this paper have been generated using the Geraldton and Perth transmissions present in the MWA data. Figure \ref{Fig:MWA_loc} shows the MWA location, the LOS receiver location (in Adelaide), the location of the two transmitters (Perth and Geraldton) and the path of the ISS during the collection period (n.b. the ISS was travelling from South West to North East relative to the MWA as indicated by the arrow).   The power of the FM transmitters is given in the parentheses next to the location of the transmitter in the figure.

In all, approximately 11 minutes of data was collected.  Unfortunately, a hardware failure occurred on one of the tiles, so 127 dual-polarised channels are used in the analysis presented here.

\section{Results}
As indicated in Section \ref{sec:exp}, preliminary analysis of the MWA data indicated the presence of the two `local'\footnote{For reference, Geraldton is approximately 294~km from the MWA and Perth is approximately 590~km distant} FM radio transmitters; Geraldton and, to a lesser extent, Perth.  In order to rule out any effects due to a mismatch in timing between the Adelaide LOS and MWA receivers, the initial analysis used the MWA data as both LOS receiver and for target surveillance. 
\subsection{First Detections}
\begin{figure}
\includegraphics[width=\columnwidth]{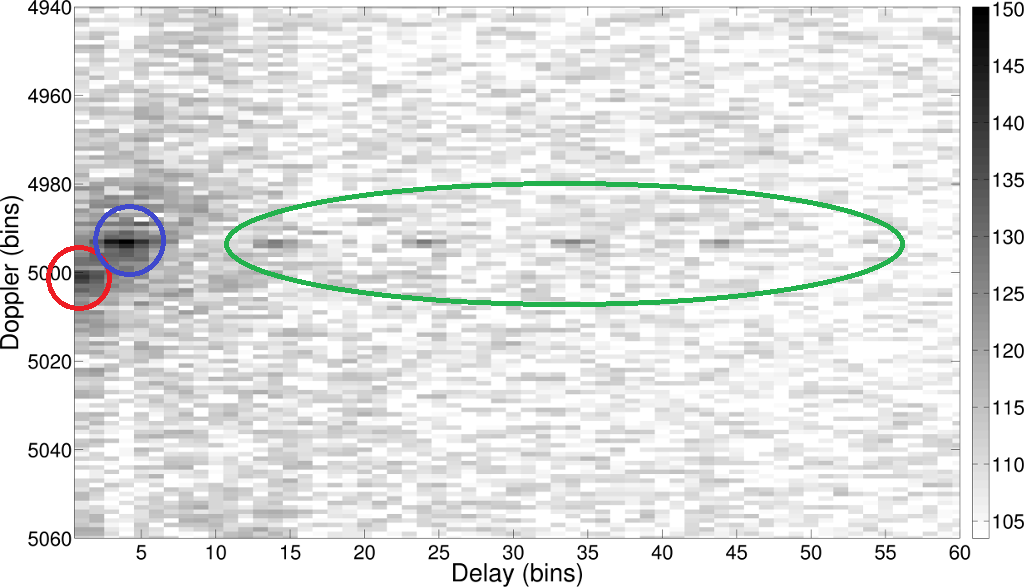}
\caption{SNR Delay Doppler map zoomed in to show the Aircraft detection (blue circle) as the large return near the zero-Doppler (red circle, bin 5,000). The aircraft shifted seven Doppler bins (corresponding to a Doppler shift of 7 Hz) and the aircraft's range sidelobes are highlighted by the green circle. Aircraft detection show on East-West polarisation using Geraldton TX with a bandwidth of 100 kHz and with a coherent processing interval (CPI) of 1 s. The intensity bar is in dB scale and represents SNR.}
\label{fig:plane_rd}
\end{figure}

After processing the data in the manner described in Section \ref{Sec:proc}, range-Doppler maps were formed that used the peak signal in the azimuth and elevation beamformed image as the surveillance channel and a beamformed reference signal directed at the Geraldton transmitter.  Example maps of the object are shown in Figure \ref{fig:plane_rd}.  Whilst the azimuth and elevation rate over time were potentially consistent with an object in orbit, the range-Doppler map clearly demonstrates that the object was not. From this figure it is evident that the object is at a range, altitude and velocity consistent with an aircraft flying at cruising altitude.  The object had a high SNR in the azimuth, elevation domain ($\approx$12~dB) and that the SNR was further enhanced by range-Doppler processing ($\approx$35~dB in Figure \ref{fig:plane_rd}).

\subsection{Meteor Detections}
\begin{figure}
\includegraphics[width=\columnwidth]{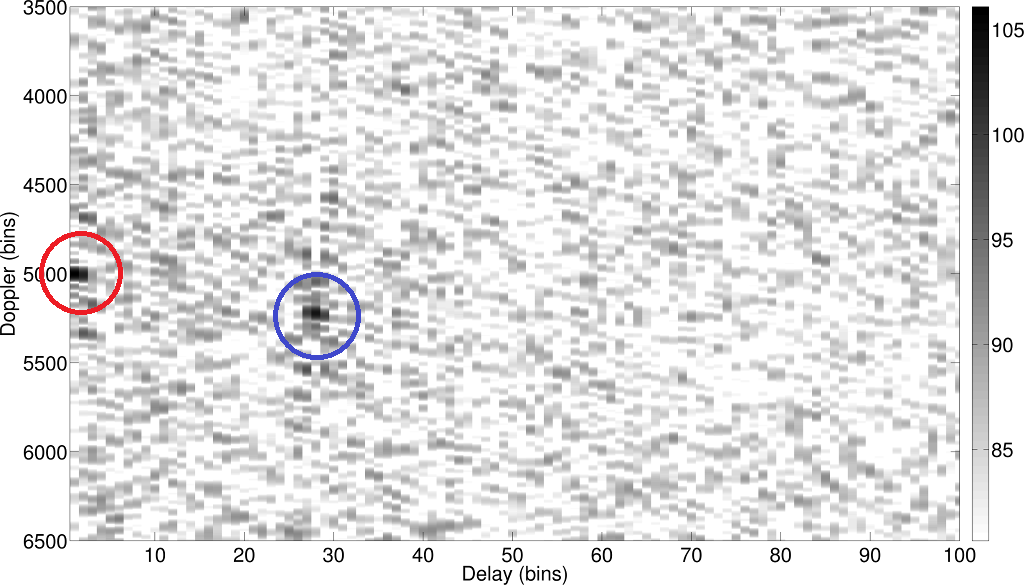}
\caption{SNR Delay Doppler map zoomed in to show the Meteor detection as the large return highlighed in blue that is significantly shifted over 200 Doppler bins - over 10,000 Hz.  Meteor detection show on East-West polarisation using Geraldton TX with a bandwidth of 100 kHz and with a CPI of 20 ms. The intensity bar is in a dB scale and represents SNR.}
\label{fig:meteor_rd}
\end{figure}

Throughout the capture, azimuth and elevation images with bright flashes were also observed.  When these bright objects were range-Doppler processed, the resultant altitudes/ranges of the objects observed were consistent with meteor returns (i.e. between 90 and 110~km).  Example provided in Figure \ref{fig:meteor_rd}.

\subsection{Detection of the ISS - Geraldton Transmitter}
\label{sub:ISS}

\begin{figure}
\includegraphics[width=\columnwidth]{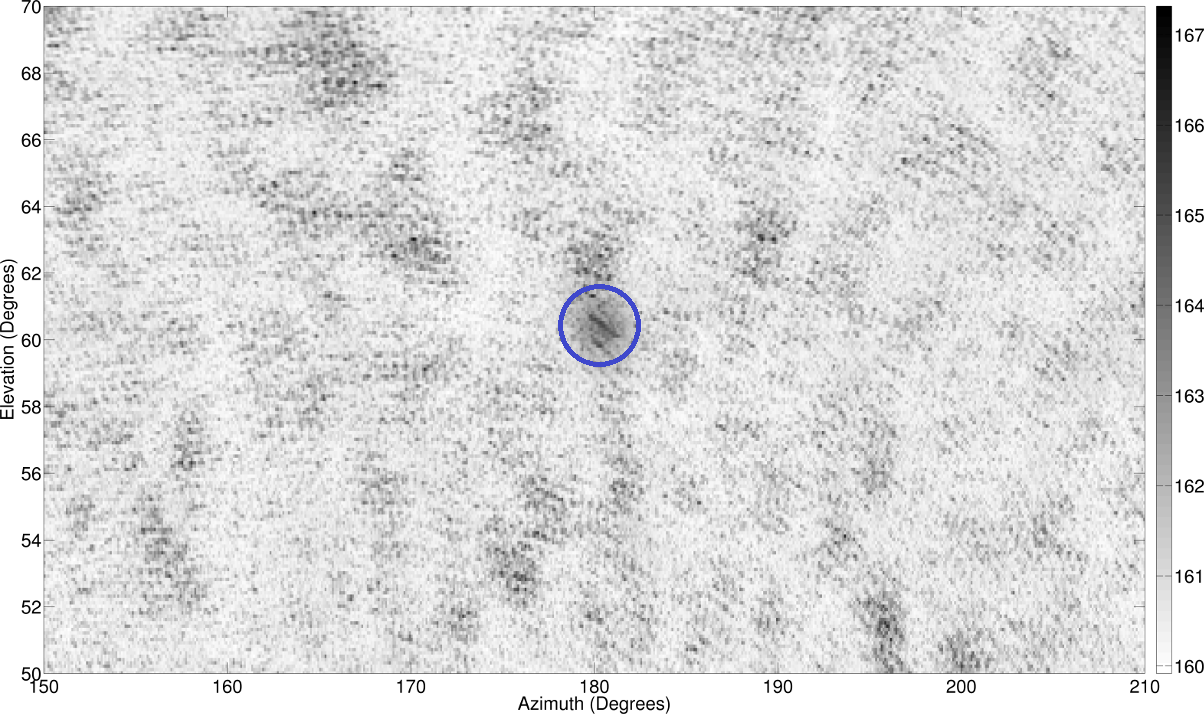}
\caption{ISS detection - highlighed in the blue circle - on the East-West polarisation using Geraldton TX - Azimuth and Elevation. The intensity bar is linear.}
\label{fig:ISS_azel}
\end{figure}

\begin{figure}
\includegraphics[width=\columnwidth]{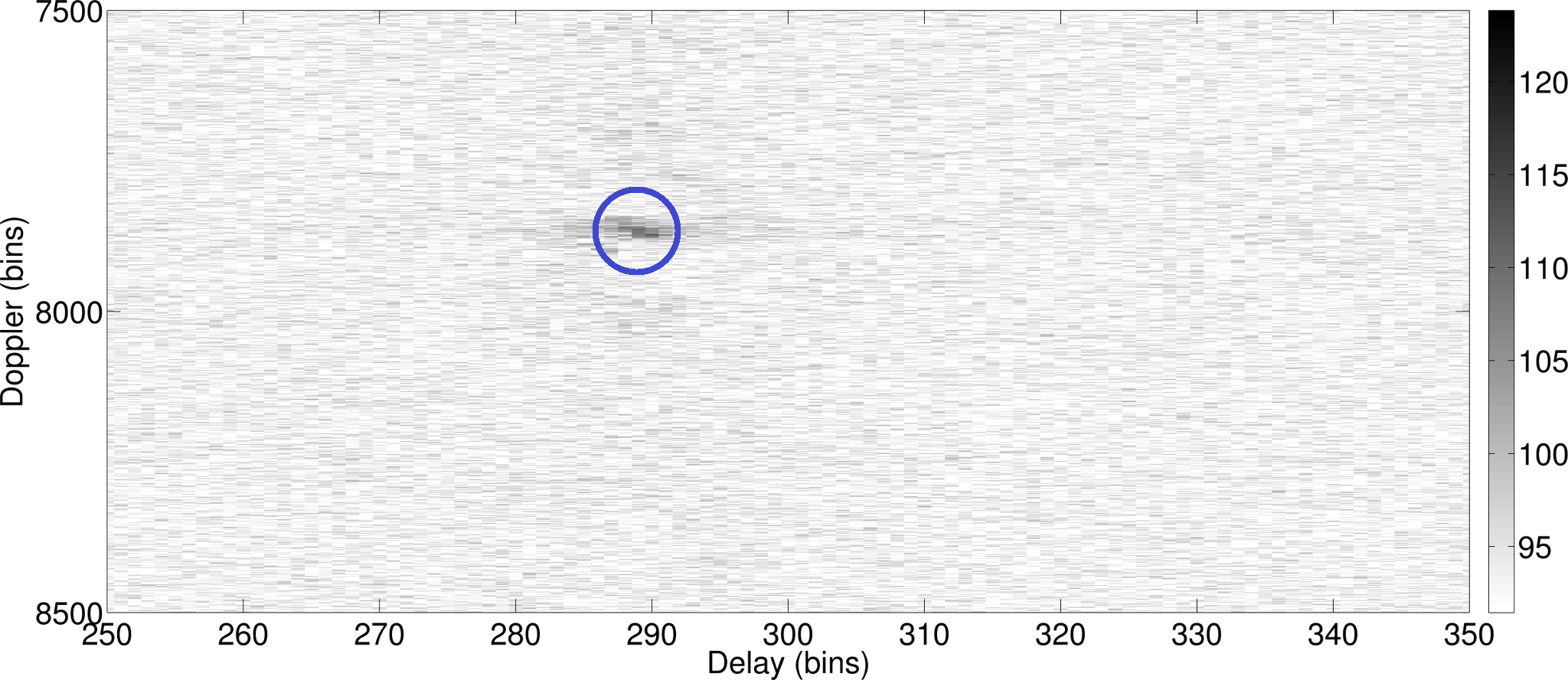}
\caption{SNR Delay Doppler map zoomed to show the ISS detection highlighted in the blue circle as the large return centered at Doppler bin 7,861, so Doppler shifted 2,861 Hz. ISS detection show on East-West polarisation using Geraldton TX with a bandwidth of 100 kHz and with a CPI of 1 s. The intensity bar is in dB scale and represents SNR.}
\label{fig:ISS_rd}
\end{figure}

Whilst beamforming a reference signal from Geraldton's transmitter, a surveillance channel beamforming process consistent with a `catalogue update' approach using azimuths and elevations calculated from published ephemeris data of the International Space Station (ISS) was employed.  An analysis of the azimuth and elevation images (an example of which is shown in Figure \ref{fig:ISS_azel}) over time indicated that the ISS was only occasionally visually discernible at the bearings indicated by the calculated values, albeit at very low SNR.  When this surveillance channel was used in forming range-Doppler maps however (an example of which is shown in Figure \ref{fig:ISS_rd}), the ISS was clearly observed (at much higher SNR) at ranges and altitudes consistent with published truth.  It is worth noting that the range-Doppler returns of the ISS were clearly discernible at detectable SNRs even when not visually discernible in the angular domain.


\subsection{Detection of the ISS - Perth Transmitter}
\begin{figure}
\includegraphics[width=\columnwidth]{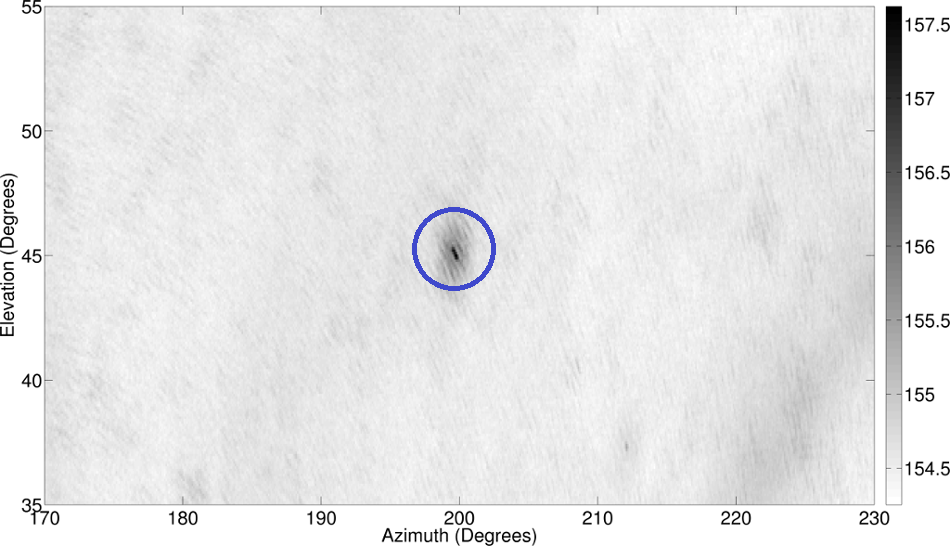}
\caption{ISS detection highlighted in the blue circle on East-West polarisation using Perth TX - Azimuth and Elevation. The colour bar is in linear scale.}
\label{fig:ISS_azel_Per}
\end{figure}

\begin{figure}
\includegraphics[width=\columnwidth]{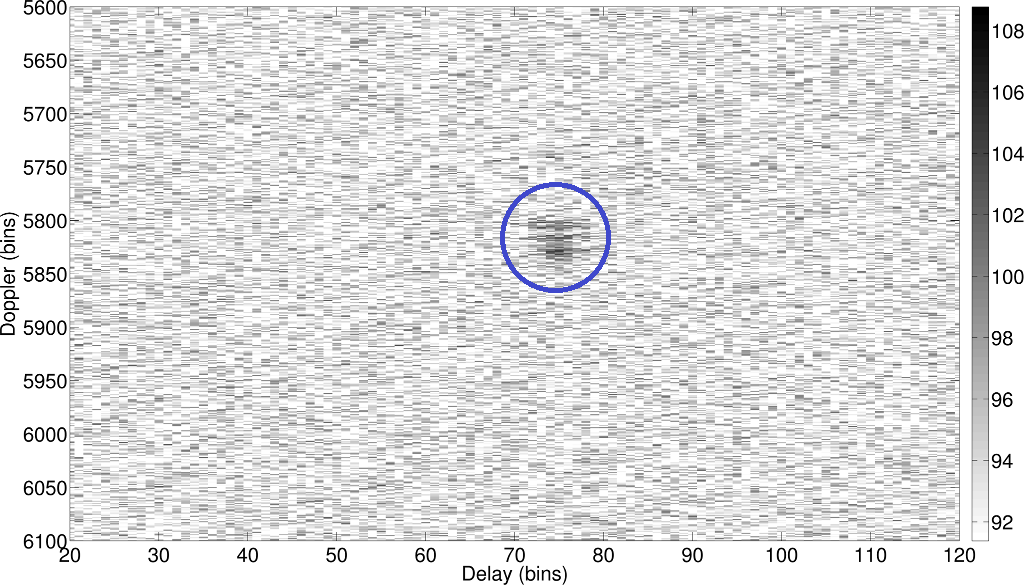}
\caption{SNR Delay Doppler map zoomed to show the ISS detection highlighted in the blue circle as the large return at Doppler bin 5,819 - shifted by 819 Hz. ISS detection show on East-West polarisation using Perth TX with a bandwidth of 50 kHz and with a CPI of 1 s. The intensity bar is in a dB scale and represents SNR.}
\label{fig:ISS_rd_Per}
\end{figure}

Figures \ref{fig:ISS_azel_Per} \& \ref{fig:ISS_rd_Per} show the same processing as Section \ref{sub:ISS} but using Perth's transmitter instead.  Whilst the SNR of the ISS in the beamformed azimuth and elevation image was typically higher than for Geraldton (the transmitter was 10x more powerful), the beamformed LOS reference was significantly weaker due to the increased ground range and, as such, the range-Doppler processing performance was significantly degraded.

The noise introduced into the processing can be substantially reduced by capturing a cleaner copy of the transmitted signal.  It is straightforward to achieve this by locating a separate time-synchronised receiver system close to the transmitter.


\subsection{Performance improvement}
\begin{figure}
\includegraphics[width=\columnwidth]{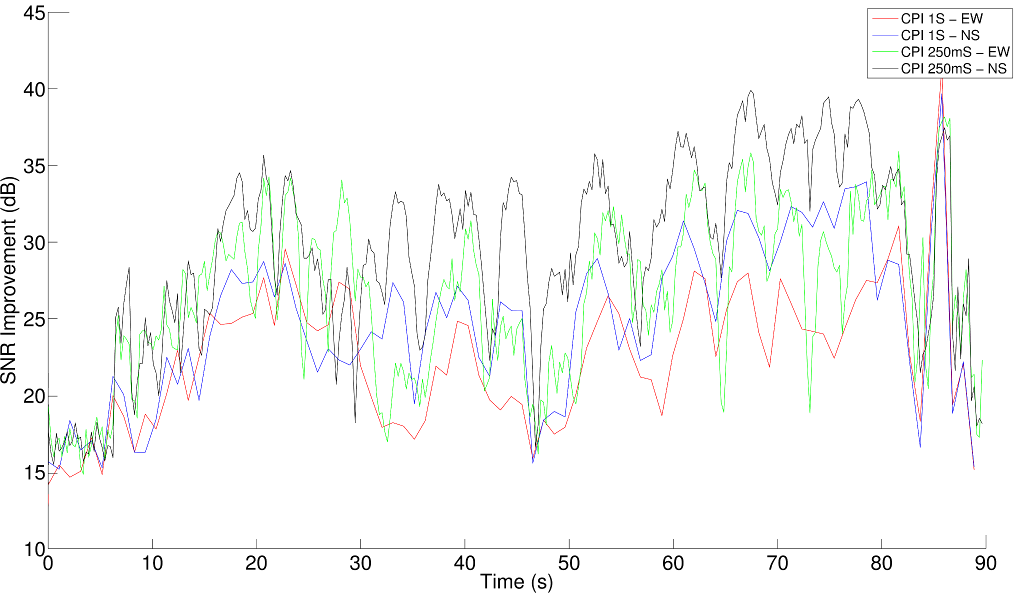}
\caption{SNR improvement achieved with range-Doppler processing}
\label{fig:SNR}
\end{figure}

Using the ISS results described in Section \ref{sub:ISS} (where a sufficient LOS-signal was available directly at the MWA), we are able to compare the SNR performance of the beamforming only approach (which is representative of a radio astronomy imaging technique) with a combined beamform and range-Doppler process. Figure \ref{fig:SNR} shows the performance improvement achieved for both polarisations and for two coherent processing intervals: 250~ms and 1~s.  It is evident from this figure, performance improvements averaging $\approx$25~dB and up to 40~dB were realised.  Only 50~kHz of bandwidth from one 10~kW station was used in this analysis.

\section{Orbit estimation}

Using the radar detections of the satellite, Bayesian statistical
methods were used for initial orbit determination.  By employing a
rigorous statistical technique, we obtain not just an estimate of the
orbit, but also the uncertainty in the orbit arising from the finite
resolution of the radar measurements.  In this case we have used an
Markov-Chain Monte-Carlo (MCMC) algorithm \cite{gamermanMCMC06} to
sample from a probability density function (pdf)
\begin{equation}
  \label{eq:3}
  p(x_0|z_{1:T}),
\end{equation}
that is the probability of the state of the system, $x_0$, at time
$t_0$, given the measurements, $z_k$, from time $t_1$ through to time
$t_T$.  This density function is defined by the state-space equations
\begin{align}
  \label{eq:1}
  x_{k+1} = f(x_k, u_k) \\
  \label{eq:2}
  z_k = h(x_k, v_k).
\end{align}
Equation~\eqref{eq:1} is the process equation that defines the
time-evolution of the state, with noise process $u_k$.
Equation~\eqref{eq:2} is the measurement function, which defines the
relationship between the state and measurement, with noise process
$v_k$.  In this case the state, $x_k$, is the 3-dimensional position
and velocity of an object in orbit and the astrodynamics of the system
are described in the process function.  The measurements that the
radar makes of the object are bistatic range, range-rate, azimuth and
elevation.  The measurement function transforms the position and
velocity of the state into radar measurement coordinates.

The orbital propagator used for the purposes of this analysis is the
standard SGP-4 propagator used by the USSTRATCOM two-line elements
(TLEs) \cite{spacetrack}.  We consider a relatively short prediction
time where we expect that the uncertainty of the orbit will be
dominated by the finite sensor resolution and so the following
analysis neglects noise in the process equation.  The measurement
noise is is defined based on features of the radar.  Noise standard
deviations used for this experiment were 0.1 degrees in azimuth, 0.2
degrees in elevation, 1 range bin and 15 Doppler bins.

\begin{figure}
  \centering
  \includegraphics[width=0.8\columnwidth]{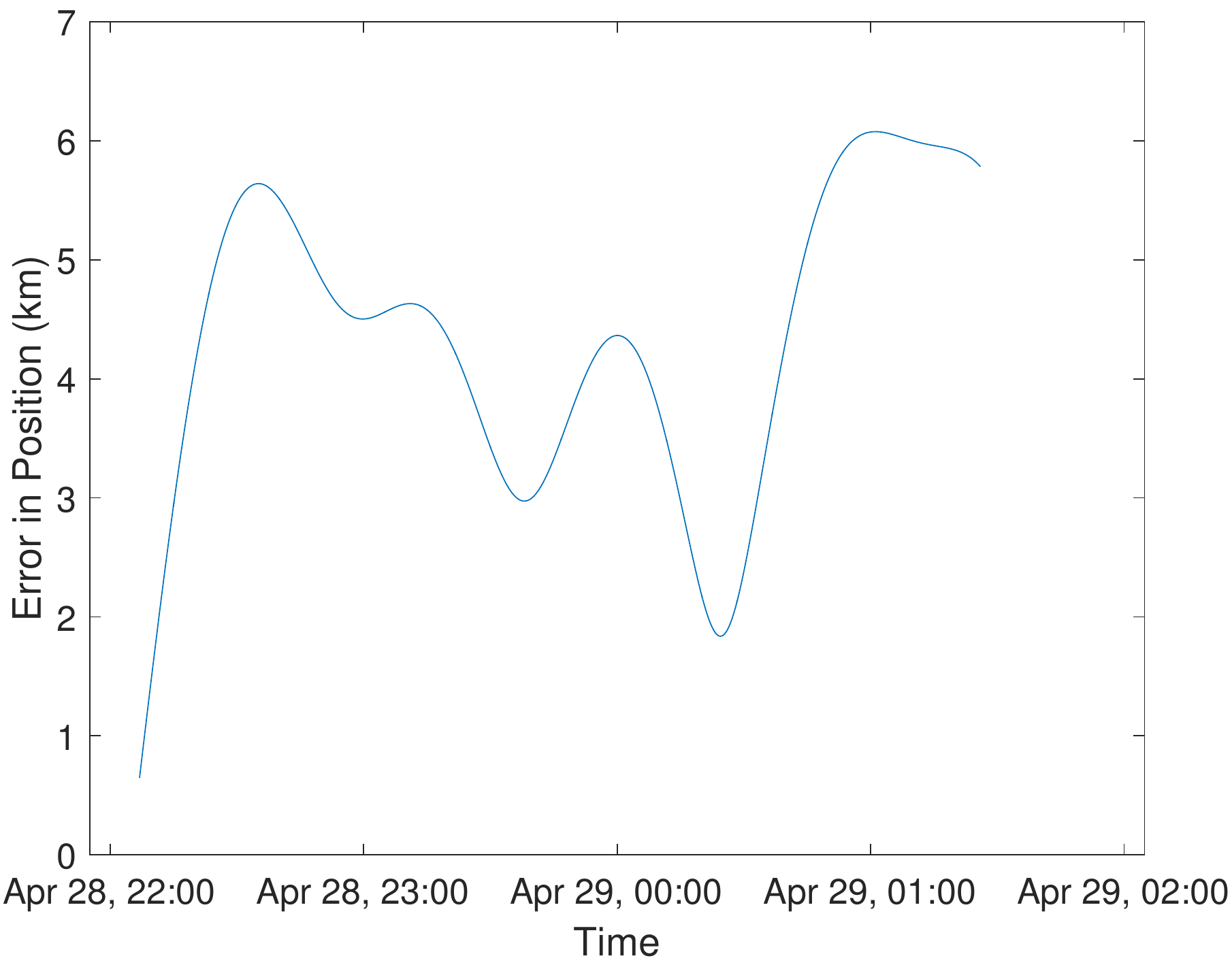}
  \caption{Difference between predictions of the orbit estimate from
    MWA data and the TLE.}
  \label{fig:pos-error}
\end{figure}

\begin{figure}
  \centering
  \includegraphics[width=0.8\columnwidth]{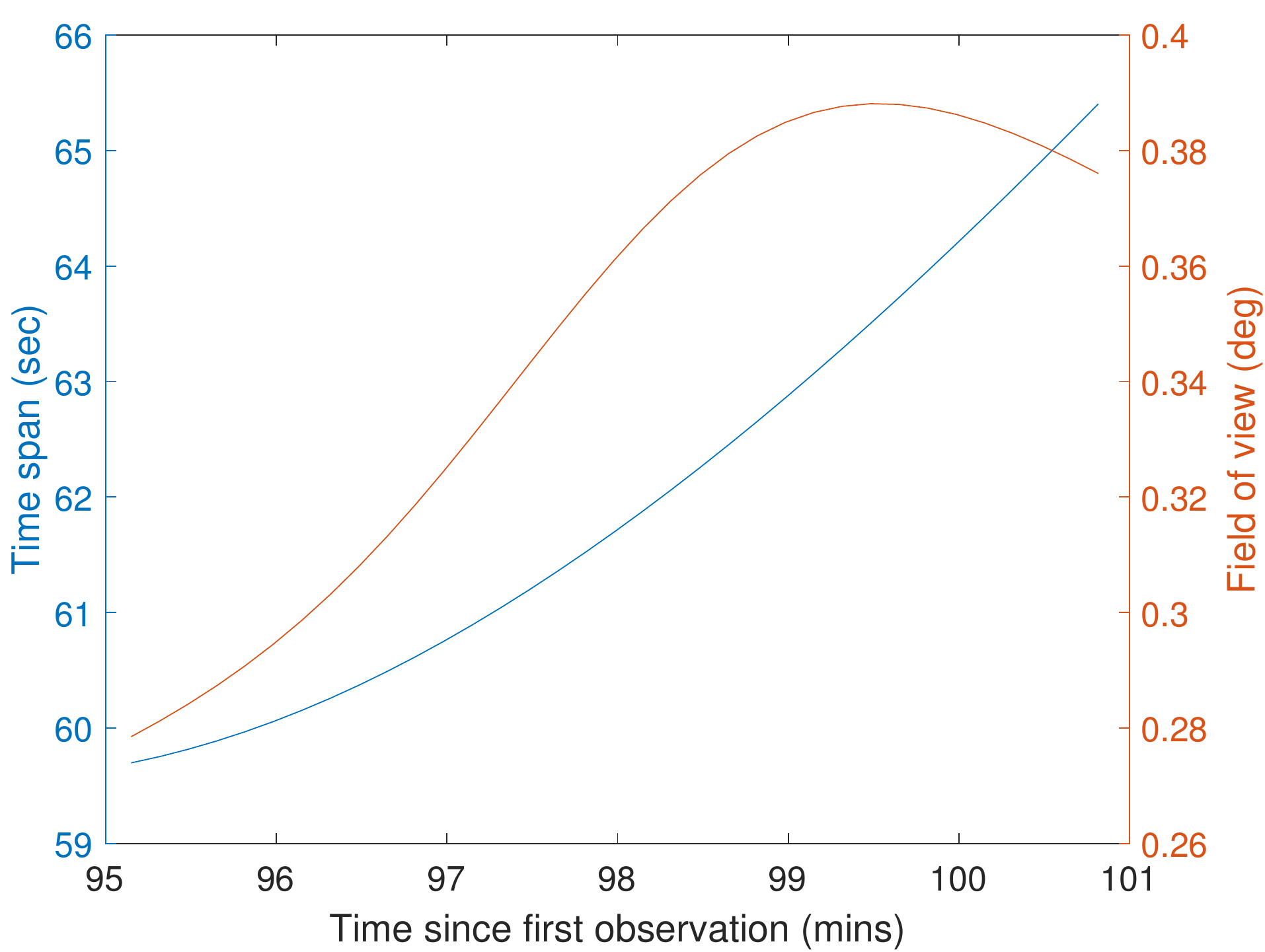}
  \caption{Difference between predictions of the orbit estimate from
    MWA data and the TLE combined with the field-of-view required to reacquire using the orbit
    estimate from the MWA data.}
  \label{fig:field-delay}
\end{figure}

In the absence of more precise ground-truth, we use the orbit
information from the TLE publicly available from USSTRATCOM.  These
are updated several times a day for the ISS and so the TLE published
closest to the MWA collection time was chosen for comparison.
Figure~\ref{fig:pos-error} shows the error in position between
predictions made from the MWA-derived orbit and the published TLE of
the ISS, using the mean of the distribution at each time.  The error
remains within several km over a three-hour prediction period.  The
uncertainty in the orbit can be split into along-track (along the path
of the orbit) and across-track (in the plane orthogonal to the path of
the orbit) errors.  The along-track error can be translated into a
timing error along the orbit, that is how early or late a sample from
our distribution might be from the ground-truth.  This is shown in
Figure~\ref{fig:field-delay}, for a sensor placed near Exmouth (shown in Figure \ref{Fig:MWA_loc}), Western
Australia, which is able to observe the transit of the ISS roughly 100
minutes after the MWA observation. The figure shows that the sensor
would need to stare for just over a minute in order to have a 99.7\%
probability of reacquiring the object on the next observable pass.
The across-track error can be translated into an effective
field-of-view required by a secondary sensor.  This is shown in
Figure~\ref{fig:field-delay}, which demonstrates that the Exmouth
sensor would require a field-of-view of less that 0.4 degrees to have
a 99.7\% probability of reacquiring the object.

\section{Future Work}
The results presented here are very preliminary and it is envisaged that significant performance improvements will be achieved through refinements in signal processing and experimental design.  More trials are being planned that will collect high SNR FM transmissions from multiple spatially distributed transmitters simultaneously with the MWA collection.  Furthermore, VHF digital TV emissions will be recorded in a dedicated collection.

\section{Conclusion}
In this paper we built upon previous work using the Murchison Widefield Array to collect reflections of terrestrial broadcast signals from objects in orbit.  The results presented here demonstrated the ability to accurately perform range-Doppler processing in a full passive radar configuration. Range-Doppler processing both provided the final two dimensions of the six dimensional state vector, and improved the detectability of the objects (through a SNR enhancement) over the previously employed angle-only approach.

\section{Acknowledgement}

This scientific work makes use of the Murchison Radio-astronomy Observatory, operated by CSIRO. We acknowledge the Wajarri Yamatji people as the traditional owners of the Observatory site. Support for the operation of the MWA is provided by the Australian Government (NCRIS), under a contract to Curtin University administered by Astronomy Australia Limited. We acknowledge the Pawsey Supercomputing Centre which is supported by the Western Australian and Australian Governments.

This research was conducted in part by the Australian Research Council Centre of Excellence for All-sky Astrophysics (CAASTRO), through project number CE110001020.

\bibliographystyle{ieeetran}
\bibliography{PassiveRadarSSA_Journal}

% Generated by IEEEtran.bst, version: 1.13 (2008/09/30)
\begin{thebibliography}{10}
\providecommand{\url}[1]{#1}
\csname url@samestyle\endcsname
\providecommand{\newblock}{\relax}
\providecommand{\bibinfo}[2]{#2}
\providecommand{\BIBentrySTDinterwordspacing}{\spaceskip=0pt\relax}
\providecommand{\BIBentryALTinterwordstretchfactor}{4}
\providecommand{\BIBentryALTinterwordspacing}{\spaceskip=\fontdimen2\font plus
\BIBentryALTinterwordstretchfactor\fontdimen3\font minus
  \fontdimen4\font\relax}
\providecommand{\BIBforeignlanguage}[2]{{%
\expandafter\ifx\csname l@#1\endcsname\relax
\typeout{** WARNING: IEEEtran.bst: No hyphenation pattern has been}%
\typeout{** loaded for the language `#1'. Using the pattern for}%
\typeout{** the default language instead.}%
\else
\language=\csname l@#1\endcsname
\fi
#2}}
\providecommand{\BIBdecl}{\relax}
\BIBdecl

\bibitem{Cherniakov08}
M.~Cherniakov, Ed., \emph{Bistatic Radars: Emerging Technology}.\hskip 1em plus
  0.5em minus 0.4em\relax Wiley, 2008.

\bibitem{NAP13456}
N.~R. Council, \emph{Continuing Kepler's Quest: Assessing Air Force Space
  Command's Astrodynamics Standards}.\hskip 1em plus 0.5em minus 0.4em\relax
  Washington, DC: The National Academies Press, 2012.

\bibitem{7024495}
N.~V. Patel, ``Averting space doom [news],'' \emph{IEEE Spectrum}, vol.~52,
  no.~2, pp. 16--17, February 2015.

\bibitem{2013PASA...30....7T}
S.~J. {Tingay}, R.~{Goeke}, J.~D. {Bowman}, D.~{Emrich}, S.~M. {Ord}, D.~A.
  {Mitchell}, M.~F. {Morales}, T.~{Booler}, B.~{Crosse}, R.~B. {Wayth}, C.~J.
  {Lonsdale}, S.~{Tremblay}, D.~{Pallot}, T.~{Colegate}, A.~{Wicenec},
  N.~{Kudryavtseva}, W.~{Arcus}, D.~{Barnes}, G.~{Bernardi}, F.~{Briggs},
  S.~{Burns}, J.~D. {Bunton}, R.~J. {Cappallo}, B.~E. {Corey}, A.~{Deshpande},
  L.~{Desouza}, B.~M. {Gaensler}, L.~J. {Greenhill}, P.~J. {Hall}, B.~J.
  {Hazelton}, D.~{Herne}, J.~N. {Hewitt}, M.~{Johnston-Hollitt}, D.~L.
  {Kaplan}, J.~C. {Kasper}, B.~B. {Kincaid}, R.~{Koenig}, E.~{Kratzenberg},
  M.~J. {Lynch}, B.~{Mckinley}, S.~R. {Mcwhirter}, E.~{Morgan}, D.~{Oberoi},
  J.~{Pathikulangara}, T.~{Prabu}, R.~A. {Remillard}, A.~E.~E. {Rogers},
  A.~{Roshi}, J.~E. {Salah}, R.~J. {Sault}, N.~{Udaya-Shankar},
  F.~{Schlagenhaufer}, K.~S. {Srivani}, J.~{Stevens}, R.~{Subrahmanyan},
  M.~{Waterson}, R.~L. {Webster}, A.~R. {Whitney}, A.~{Williams}, C.~L.
  {Williams}, and J.~S.~B. {Wyithe}, ``{The Murchison Widefield Array: The
  Square Kilometre Array Precursor at Low Radio Frequencies},''
  \emph{Publications of the Astronomical Society of Australia}, vol.~30, p.
  e007, Jan. 2013.

\bibitem{2013AJ....145...23M}
B.~{McKinley}, F.~{Briggs}, D.~L. {Kaplan}, L.~J. {Greenhill}, G.~{Bernardi},
  J.~D. {Bowman}, A.~{de Oliveira-Costa}, S.~J. {Tingay}, B.~M. {Gaensler},
  D.~{Oberoi}, M.~{Johnston-Hollitt}, W.~{Arcus}, D.~{Barnes}, J.~D. {Bunton},
  R.~J. {Cappallo}, B.~E. {Corey}, A.~{Deshpande}, L.~{deSouza}, D.~{Emrich},
  R.~{Goeke}, B.~J. {Hazelton}, D.~{Herne}, J.~N. {Hewitt}, J.~C. {Kasper},
  B.~B. {Kincaid}, R.~{Koenig}, E.~{Kratzenberg}, C.~J. {Lonsdale}, M.~J.
  {Lynch}, S.~R. {McWhirter}, D.~A. {Mitchell}, M.~F. {Morales}, E.~{Morgan},
  S.~M. {Ord}, J.~{Pathikulangara}, T.~{Prabu}, R.~A. {Remillard}, A.~E.~E.
  {Rogers}, A.~{Roshi}, J.~E. {Salah}, R.~J. {Sault}, N.~{Udaya Shankar}, K.~S.
  {Srivani}, J.~{Stevens}, R.~{Subrahmanyan}, R.~B. {Wayth}, M.~{Waterson},
  R.~L. {Webster}, A.~R. {Whitney}, A.~{Williams}, C.~L. {Williams}, and
  J.~S.~B. {Wyithe}, ``{Low-frequency Observations of the Moon with the
  Murchison Widefield Array},'' \emph{The Astronomical Journal}, vol. 145,
  p.~23, Jan. 2013.

\bibitem{2013AJ....146..103T}
S.~J. {Tingay}, D.~L. {Kaplan}, B.~{McKinley}, F.~{Briggs}, R.~B. {Wayth},
  N.~{Hurley-Walker}, J.~{Kennewell}, C.~{Smith}, K.~{Zhang}, W.~{Arcus},
  N.~D.~R. {Bhat}, D.~{Emrich}, D.~{Herne}, N.~{Kudryavtseva}, M.~{Lynch},
  S.~M. {Ord}, M.~{Waterson}, D.~G. {Barnes}, M.~{Bell}, B.~M. {Gaensler},
  E.~{Lenc}, G.~{Bernardi}, L.~J. {Greenhill}, J.~C. {Kasper}, J.~D. {Bowman},
  D.~{Jacobs}, J.~D. {Bunton}, L.~{deSouza}, R.~{Koenig}, J.~{Pathikulangara},
  J.~{Stevens}, R.~J. {Cappallo}, B.~E. {Corey}, B.~B. {Kincaid},
  E.~{Kratzenberg}, C.~J. {Lonsdale}, S.~R. {McWhirter}, A.~E.~E. {Rogers},
  J.~E. {Salah}, A.~R. {Whitney}, A.~{Deshpande}, T.~{Prabu}, N.~{Udaya
  Shankar}, K.~S. {Srivani}, R.~{Subrahmanyan}, A.~{Ewall-Wice}, L.~{Feng},
  R.~{Goeke}, E.~{Morgan}, R.~A. {Remillard}, C.~L. {Williams}, B.~J.
  {Hazelton}, M.~F. {Morales}, M.~{Johnston-Hollitt}, D.~A. {Mitchell},
  P.~{Procopio}, J.~{Riding}, R.~L. {Webster}, J.~S.~B. {Wyithe}, D.~{Oberoi},
  A.~{Roshi}, R.~J. {Sault}, and A.~{Williams}, ``{On the Detection and
  Tracking of Space Debris Using the Murchison Widefield Array. I. Simulations
  and Test Observations Demonstrate Feasibility},'' \emph{The Astronomical
  Journal}, vol. 146, p. 103, Oct. 2013.

\bibitem{5136190}
P.~E. Dewdney, P.~J. Hall, R.~T. Schilizzi, and T.~J. L.~W. Lazio, ``The square
  kilometre array,'' \emph{Proceedings of the IEEE}, vol.~97, no.~8, pp.
  1482--1496, Aug 2009.

\bibitem{2015PASA...32....5T}
S.~E. {Tremblay}, S.~M. {Ord}, N.~D.~R. {Bhat}, S.~J. {Tingay}, B.~{Crosse},
  D.~{Pallot}, S.~I. {Oronsaye}, G.~{Bernardi}, J.~D. {Bowman}, F.~{Briggs},
  R.~J. {Cappallo}, B.~E. {Corey}, A.~A. {Deshpande}, D.~{Emrich}, R.~{Goeke},
  L.~J. {Greenhill}, B.~J. {Hazelton}, M.~{Johnston-Hollitt}, D.~L. {Kaplan},
  J.~C. {Kasper}, E.~{Kratzenberg}, C.~J. {Lonsdale}, M.~J. {Lynch}, S.~R.
  {McWhirter}, D.~A. {Mitchell}, M.~F. {Morales}, E.~{Morgan}, D.~{Oberoi},
  T.~{Prabu}, A.~E.~E. {Rogers}, A.~{Roshi}, N.~{Udaya Shankar}, K.~S.
  {Srivani}, R.~{Subrahmanyan}, M.~{Waterson}, R.~B. {Wayth}, R.~L. {Webster},
  A.~R. {Whitney}, A.~{Williams}, and C.~L. {Williams}, ``{The High Time and
  Frequency Resolution Capabilities of the Murchison Widefield Array},''
  \emph{Astronomical Society of Australia}, vol.~32, p. e005, Feb. 2015.

\bibitem{4654011}
J.~Palmer, D.~Merrett, S.~Palumbo, J.~Piyaratna, S.~Capon, and H.~Hansen,
  ``Illuminator of opportunity bistatic radar research at dsto,'' in \emph{2008
  International Conference on Radar}, Sept 2008, pp. 701--705.

\bibitem{gamermanMCMC06}
D.~Gamerman and H.~F. Lopes, \emph{Markov Chain Monte Carlo: Stochastic
  Simulation for Bayesian Inference}.\hskip 1em plus 0.5em minus 0.4em\relax
  Taylor and Francis CRC Press, 2006.

\bibitem{spacetrack}
USSTRATCOM, ``Space-track,'' Online, \url{space-track.org}.

\end{thebibliography}

\end{document}